# Multiple Imputation with Massive Data: An Application to the Panel Study of Income Dynamics[*]

Yajuan Si, Steve Heeringa, David Johnson, Roderick J. A. Little, Wenshuo Liu, Fabian Pfeffer,

and Trivellore Raghunathan

Yajuan Si is Research Assistant Professor, Survey Research Center, Institute for Social Research, University of Michigan, Ann Arbor, MI, USA (Email: yajuan@umich.edu).

Steve Heeringa is Senior Research Scientist, Survey Research Center, Institute for Social Research, University of Michigan, Ann Arbor, MI, USA.

David Johnson is Research Professor, Survey Research Center, Institute for Social Research, University of Michigan, Ann Arbor, MI, USA.

Roderick J. A. Little is Professor, Department of Biostatistics, School of Public Health; Research Professor, Survey Research Center, Institute for Social Research, University of Michigan, Ann Arbor, MI, USA.

Wenshuo Liu is Senior Research Scientist, Research and Innovation, Interactions LLC.

Fabian Pfeffer is Associate Professor, Department of Sociology; Research Associate Professor, Survey Research Center, Institute for Social Research, University of Michigan, Ann Arbor, MI, USA.

Trivellore Raghunathan is Professor, Department of Biostatistics, School of Public Health; Research Professor, Survey Research Center, Institute for Social Research, University of Michigan, Ann Arbor, MI, USA

---

[*] This work was supported by the National Institute on Aging of the National Institutes of Health [R01AG040213].

**Abstract**

Multiple imputation (MI) is a popular and well-established method for handling missing data in multivariate data sets, but its practicality for use in massive and complex data sets has been questioned. One such data set is the Panel Study of Income Dynamics (PSID), a longstanding and extensive survey of household income and wealth in the U.S. Missing data for this survey are currently handled using traditional hot deck methods because of the simple implementation; however, the univariate hot deck results in large random wealth fluctuations. MI is effective but faced with operational challenges. We use a sequential regression/ chained-equation approach, using the software IVEware, to multiply impute cross-sectional wealth data in the 2013 PSID, and compare analyses of the resulting imputed data with those from the current hot deck approach. Practical difficulties, such as non-normally distributed variables, skip patterns, categorical variables with many levels, and multicollinearity, are described together with our approaches to overcoming them. We evaluate the imputation quality and validity with internal diagnostics and external benchmarking data. MI produces improvements over the existing hot deck approach by helping preserve correlation structures, such as the associations between PSID wealth components and the relationships between the household net worth and socio-demographic factors, and facilitates completed data analyses with general purposes. MI incorporates highly predictive covariates into imputation models and increases efficiency. We recommend the practical implementation of MI and expect greater gains when the fraction of missing information is large.

Abstract word count: 240; Main text word count: 5734.

*Statement of Significance:* In this paper, we address the challenges and strategies in implement multiple imputation (MI) in a practical setting. MI is now a popular and well-established method for handling missing data; however, misunderstandings of the method exist mainly due to its practicality for use in massive data. Motivated by the Panel Study of Income Dynamics, which provides rich survey data on household income and wealth in the U. S., we demonstrate our strategies handling the practical difficulties of MI in the wealth component imputation, to offer detailed advice for general researchers interested in applying MI with a complex dataset.

## 1. Introduction

Multiple imputation (MI) is a useful tool for dealing with missing data, given its attractive theoretical properties, its ability to handle any pattern of missing data, and the numerous computation platforms that are available in practice. Since the initial development by Rubin (1987), MI has been successfully applied in a variety of fields for missing data and more broadly to handle related problems such as measurement error, confidentiality protection, and finite population inference (Reiter and Raghunathan, 2007; Van Buuren, 2012; Carpenter and Kenward, 2013).

The big data era has led to the increased availability of massive datasets. Here we focused on massive data that have many variables, different variable types and distributions, skipped patterns and a complex dependency structure. Loh et al. (2019) argued that MI under parametric models could not be successfully implemented in these settings. Others (e.g., Little, 2020) contested this assertion, and Stuart et al. (2009) and He et al. (2010) successfully applied chained-equation approaches to large data sets. This paper aimed to illustrate some challenges in implementing MI on a large dataset by imputing an extensive set of variables in the Panel Study of Income Dynamics (PSID). The goal of the project was to multiply-impute missing data in PSID's 18 wealth components, along with the missing values of predictors of these components, for public data release. We considered a total of 409 variables with varying amounts of missing data. We note that our application was considerably more complex than that considered by Loh et al. (2019), who restricted attention to the imputation of a single variable (amount of interest and dividend income) in the U.S. Consumer Expenditure Survey.

The current approach implemented by PSID was a univariate hot deck imputation, where each case of item nonresponse (and bracketed response) was flagged and assigned a value that was randomly drawn from the set of reported values (or within the same bracket) with selection probabilities equal to the distribution of observed continuous values (or within the respective bracket) (for details see Pfeffer and Griffin, 2017). This approach has three fundamental limitations: the current hot deck approach does not condition on covariate information, it treats each source of wealth independently, and it does not allow the user to incorporate imputation uncertainty into estimates of standard errors. Moreover, even a small percentage of the hot deck imputations results in large random wealth fluctuations between waves. Pfeffer and Griffin (2017) have found that real changes in life circumstances often account for large changes in wealth, so using MI to include additional information in the imputation will help reduce the random variation. Some form of multivariate hot deck within adjustment cells would allow for the incorporation of some associations between variables (see, e.g., Haziza and Beaumont, 2007, Andridge and Little, 2010), but the method is ill-suited to handling the complex multivariate pattern of missing data, the fact that different covariates may be predictive of each incomplete variable, and the need to simultaneously reflect the relationships between a large number of variables. Our goal is to address these issues by applying MI, including as predictors various socioeconomic and demographic characteristics, the values of income and other asset components, and known predictors of wealth fluctuations such as changes in household composition, employment and retirement status, health conditions, and residence.

We present our solutions to various practical difficulties that arise in this and similar big-data applications. We also evaluate the imputation quality and validity with internal diagnostics and external benchmarking data, and demonstrate the improvements MI produces over the existing hot deck approach. The paper structure is organized as follows. Section 2 reviews the background of MI methodology and existing software for both joint and sequential regression imputations. Section 3 describes the motivating PSID study with general issues related to applications with incomplete massive datasets. We depict our imputation approaches and evaluation criteria in Section 4 and illustrate the results of the PSID application in Section 5. Section 6 concludes with discussions of the main findings and extensions.

## 2. Background

MI fits models to predict the missing values with predictive covariates based on the observed values (Little, Carpenter, and Lee, 2021). Grounded in Bayesian methodology, MI draws model parameters from their posterior distributions and then imputes draws from a posterior predictive distribution or using predictive mean matching. The result is $M$ completed datasets[†], each with different draws or imputations of the missing values. Variance estimation combines the within-imputation and between-imputation variance across these $M$ datasets, using simple MI combining rules (Rubin, 1987). Although its etiology is Bayesian, MI has been shown to yield efficient estimates and inferential validity from the frequentist perspective. The proper MI model should be general enough that covers all potential analyses of interest to yield congeniality (Rubin, 1996; Meng, 2002).

Various MI software has been developed based on joint multivariate normal distributions, e.g., PROC MI (SAS Institute Inc., 2017) in SAS, the suite of mi commands and ice in Stata (StataCorp., 2021), Amelia (King et al., 2001; available as an independent program and an R package) and the R package norm (Schafer, 1997), or a sequence of fully conditional distributions, e.g., IVEware (Raghunathan et al., 2001; available with multiple interfaces), the R package mice (Van Buuren and Oudshoorn, 1999), and the R package mi (Gelman et al., 2015). MI for multilevel models is available in the R packages jomo (Quartagno, Grund, and Carpenter, 2019) and pan (Schafer, 2016). Flexible nonparametric Bayesian mixture models are also applied to jointly impute many incomplete categorical variables and a mixed group of categorical and continuous variables, e.g., NPBayesImputeCat (Si and Reiter, 2013) and MixedDataImpute (Murray and Reiter, 2016). Other prediction algorithms applied to chained equation MI include classification and regression trees (Burgette and Reiter, 2010), Bayesian additive regression trees (Xu et al., 2016), and random forests (Stekhoven and Bühlmann, 2012).

---

[†] We use "imputed data" to represent the imputed values of missing data, use "completed data" to represent the combination of observed data and imputed data, and use "complete data" to represent observed data.

With a large number of incomplete variables of mixed types and with various structural restrictions, developing a coherent joint imputation model is challenging. In contrast, chained equation or sequential regression imputation approaches are both flexible and computationally attractive. This approach models the regression of each variable with missing values on all the other variables in the data set, varying the type of regression models by the type of variables being imputed. Covariates include all other variables observed or imputed for that individual. The sequence of imputing missing values overwrites previously drawn values, building interdependence among imputed values and exploiting the correlational structure among covariates. It cycles iteratively through the dataset, imputing the missing values of each variable in turn. The cycles are similar to those of a Bayesian Gibbs' sampler, but the method is only approximately Bayesian because the sequence of conditional distributions may not correspond to a coherent joint distribution (Zhu and Raghunathan, 2005; Liu et al., 2014).

Previous studies have applied MI to large data sets with hundreds of variables and discussed detailed steps in the implementation (e.g., Stuart et al., 2009; He et al., 2010; Azur et al., 2011; and Drechsler, 2011). We work with a similar data setting but develop a systematic process with step-by-step solutions to a broad list of problems often encountered in practice, such as data transformations, variable selection, restrictions, and diagnostics. Furthermore, we supplement the previous studies by comparing MI to hot deck imputations, examining the fraction of missing information, and conducting external evaluations.

We carry out chained equation MI using IVEware (Raghunathan, 2020), which can automatically handle issues of structural zeros, restrictions, and bounded values that are present in our application, and also has options for variable selection that are important given our large set of potential predictors. Drechsler (2011) summarizes features included in different software packages. Not all these capabilities are currently available in alternative chained equation software. For example, the R packages mice and mi cannot directly handle skip patterns and require additional programming efforts. The IVEware imputations order variables by the amount of missing values from least to most, and draw from the posterior predictive distribution specified by the regression model with a flat or relatively noninformative prior distribution for the parameters in the regression model. Informative prior distributions can be introduced to the imputation model in mi and facilitate the variable selection. However, the PROC MI and the mi suite in Stata do not have the capabilities mentioned above.

### 3. Motivating case study

The PSID began in 1968 with a sample of over 18,000 individuals living in 5,000 U. S. families, and it has followed them for the past five decades to study the U.S. population (PSID, 2020). The PSID sample is dynamic and grows as children and grandchildren from these families form their own households and are recruited into the PSID sample and longitudinal data collection. One of the key study topics of interest to researchers is the collection of wealth information for the households of these individuals and their descendants. The survey instrument contains questions on a range of separate wealth components, such as home values, mortgages, different types of financial assets, real assets, and debts; together, these components form a measure of net worth. Item nonresponse for most wealth components is quite low in the PSID (< 5%) partly due to the study's use of unfolding brackets and editing system to minimize nonresponse. However, about 20%–25% of families do not report a continuous value (e.g., either report a bracketed value or no value) for at least one of the components needed to compute net worth. Consequently, since the first wealth module in 1984, the PSID has imputed missing values for users and summed the imputed components to create aggregate net worth measures.

We focused on cross-sectional PSID data for 2013, including all families who responded (9,063) and 409 variables that we selected to capture key socio-demographic characteristics of households, household reference persons, and partners, such as employment, wages, family income, consumption, race, education, and wealth. We drew on family-level and individual-level information as potential predictors of wealth components. The family-level predictors included income, consumption, and other financial measures, and the individual-level predictors included socio-demographic variables, employment, wage, and income information for household reference persons and partners (for a full list, see the Supplement; PSID, 2013). As part of the chained equation regressions, all components of wealth also became predictors of other components of wealth.

In datasets with structural zeros, where certain variables are "not applicable" given values of other variables, it is important to code variables in a way that distinguishes between "not applicable" and "missing." For instance, in the case of wealth variables, PSID includes filter questions that ask whether a household holds a certain asset or not, such as whether the household owns their home. To those indicating homeownership, a follow-up question is fielded to ascertain the value of the home and the remaining mortgages. For those indicating no homeownership, the follow-up variable is recorded as "non-applicable". Both the house ownership indicator and the house value variable may be missing.

## 4. Methods and evaluation criteria

We now describe the process of creating MI datasets based on sequential regression imputation models, from data preparation, developing imputation models, creating multiple imputations, to model diagnostics.

### 4.1 Data preparation

**Sample design and weights.** The design information (strata, clustering, and weights) may be informative for the imputation process. Methods for handling clustering variables and survey weights include models with random effects (Reiter et al., 2006; Quartagno, Carpenter, and Goldstein, 2020), and penalized splines of propensity prediction (Zheng and Little, 2003). IVEware has implemented the weighted finite population Bayesian bootstrap methods to incorporate weighting, clustering, and stratification in the synthetic population generation (Zhou, Elliott, and Raghunathan, 2016). However, the computation with large-scale datasets is challenging. The sequential imputation model in IVEware provides flexibility in the selection of covariates, and we assume that the design information becomes noninformative after conditioning on the variables that are highly predictive of wealth. IVEware is yet unable to account for clustering effects and fit multilevel models. The imputation model specification and sequential implementation need novel modifications to properly account for the design information. We thus omitted the design information in the imputation process and considered it as a possible future extension in the Discussion.

**Recoding missing values.** We imputed missing values for all wealth measures and predictors. As mentioned above, different values (e.g., 999999, 0 or NA) indicating missing data are recoded as the same missing flags. PSID also provides flags for some variables to indicate changes to the originally reported value that resulted from editing or other processing steps. Cases that are edited during data cleaning are not considered missing. If the flag code showed that the value had been imputed by other methods, we recoded it as missing. The "not applicable" values were treated as such and passed over in the imputation process via specified restrictions.

**Outlier detection**. Typically, outlier detection has been carried out in the data cleaning process to detect errors with pre-specified editing rules. There are also some legitimate observations that are not errors and may be extreme and could be influential. Influential units can be handled with robust methods (e.g., Chen, Haziza, and Michal, 2020). We plotted the frequency distributions of observed values to detect extreme values that would introduce skewness to the distributions of the individual variables.

**Transformations.** With legitimately extreme values, we performed transformations before imputation to limit the influence of outliers. After visually checking the frequency histograms of the observed values and calculating skewness parameters, we chose the cube root transformation for the wealth, income, and wage variables. This transformation substantially symmetrizes the shape of the distribution. Unlike the logarithm, the cube root transformation can be applied to negative and zero values, which occur for some variables in our study. Congeniality between the imputation model and the analysis model is an important consideration, requiring that the imputation of inclusive of variables that could be included in an analysis model and that are associated with the outcome (Xie and Meng, 2017).

## 4.2. Developing the imputation models

In fitting the imputation models, multicollinearity issues were the main source of run errors. One reason was that PSID supplemented with slightly different recoded versions of the same variables as well as versions of aggregated variables created based on user interests. Naturally, the aggregated variables are collinear with their components. Before the imputation, we applied principal component analyses to identify such variables and remove them from the set of predictors for the imputation models.

Another cause of multicollinearity was categorical variables with many nominal levels, such as the state of residence with 51 values. This variable was included as a predictor in the regression model with 51 dummy variables, leading to potential collinearity, especially from categories with small sample sizes, in this case, small states. Before the imputation process, we used forward selection to select the dummy variables along with other predictors to avoid issues with collinearity. Case identifiers, flag variables, and boundaries of intervals were not used as predictors during imputation but used in defining whether the imputation should be done or not and for the boundaries for the imputed values.

Imputations are draws from the predictive posterior distribution of each missing variable, based on a regression model for each incomplete variable, preferably with all the predictors. Given a large number of predictors, we used forward variable selection at each imputation cycle to identify a subset of the predictors tailored to the variable type: specifically, linear for continuous variables, logistic or multinomial logistic for categorical variables, and Poisson for count variables. For semi-continuous variables, such as wealth, income, and consumption, a two-stage model was used to impute missing values of a mix of a binary variable indicating presence or absence and a continuous value if the variable is present. First, a logistic regression model was used to impute the presence or absence of an asset. Conditional on imputing a non-zero status, a normal linear regression model for the cube-root transformed outcome value was then used to impute non-zero values. For example, we first imputed the indicator of whether the family had any real estate and then imputed the real estate value if owned. The indicators of non-zero status and amounts if non-zero then became potential predictors in imputation models for other variables.

Eliminating covariates that were not predictive of the outcome variable with missing values avoided problems with multicollinearity and helped to improve the convergence of the IVEware iterations. At each imputation cycle, IVEware has the ability to use the marginal increase of the goodness of fit statistic $R^2$ when including each variable to select the variable. We set a minimum $R^2$ increase of 0.005 and a maximum number of predictors of 10. If the categorical variable is declared as a single variable, the variable with multiple categories is jointly tested, excluded, or included. If dummy variables are created and listed separately, each individual dummy variable is tested. The imputation model selects variables strongly correlated with the outcome and improves efficiency; inclusion of variables not related to the outcome will inflate variances (Little and Vartivarian, 2005).

The forward variable selection method to determine the models is admittedly rather ad hoc, but perhaps justified given the size and complexity of the problem, the need to avoid collinearities, and the goal of prediction of the missing values rather than interpretation as in a substantive model. Hot deck imputation with appropriately defined adjustment cells is a possible alternative, but regression-based methods are much more flexible about conditioning on an extensive set of covariates than the hot dock. The hot deck has an implicit model that, like any model, needs to be checked (David et al., 1986). In principle, more sophisticated methods such as ridge regression or lasso could be implemented (see, e.g., Deng et al., 2016), although doing so would be challenging in this particular missing data setting. Often additive models are the default option in regression models, but interactions that are potentially predictive can and should be explicitly included as covariates. We can use tree-based methods that can detect interactions and non-linear relationships as imputation approaches. IVEware needs extensions to allow rigorous variable selection and flexible imputation algorithms.

All regression models were fitted only to the set of applicable cases; for example, the regression model for the imputation of house values was restricted to households that own their homes. Because homeownership can itself be imputed, the set of applicable cases changes in each imputation iteration. These restrictions can be nested and must be explicitly specified so that the higher-level restricting variables are not used as predictors in the regression model. Restrictions also arise from nested skip patterns in the questionnaire. For example, a question about a loan for a second house is asked only when the respondent indicates first having a house and then a second house. Given specified restrictions, some variables are constrained to be positive, for example, the value of an owned house. However, other variables could still take on 0 values with restrictions, requiring a semi-continuous variable declaration. For example, a family could own real estate (restriction) but with or without a mortgage, where the mortgage value is a semi-continuous variable.

Some missing values come with logical or consistency bounds that must be accounted for during imputation. For example, the annual property tax or insurance premium amount must be non-negative. For some survey variables such as wealth and income, some respondents did not provide an exact value and instead answered follow-up questions asking for brackets or range for survey variable, which then defined bounds within which the imputed values must lie. The bounds are incorporated by drawing imputations from a predictive distribution restricted to lie within the bounds, which is an option in IVEware.

### 4.3. Creating imputations from IVEware multiple chains

The number of cycles in the initial burn-in period, the number of iterations between creating one set of imputations, and the number $M$ of imputations to be performed, together determine the total number of cycles of the Gibbs' type algorithm.

### 4.4. Model diagnostics

It is important to check that imputations are plausible as unanticipated problems in setting up the regressions can lead to poor imputations. Imputed and observed values can be compared using graphical and numeric diagnostic tools (Stuart et al., 2009). The marginal distributions of observed and imputed values are expected to be similar under missing completely at random but may markedly differ under missing at random conditions. Nevertheless, comparisons across the imputed data sets will be useful as the first phase of evaluation. Bondarenko and Raghunathan (2016) developed graphical and numeric diagnostic tools for MI to compare the distributions of imputed and observed values conditional on the response propensity score. However, such tools need the extension to work for imputation with restrictions and semi-continuous variables. A useful feature of MI is that the fraction of missing information (FMI), which estimates the relative increase in variance due to missing data (Rubin, 1987; Raghunathan, 2016), is readily computed as a simple function of the between-imputation and within-imputation variance. Note that no comparable measure is available from a single imputation method. A large value of FMI indicates a substantial increase in variance due to nonresponse and a high level of uncertainty about the imputation process.

The current hot deck imputation method performs univariate imputations, ignoring relationships with other variables, although other hot deck imputation approaches within adjustment cells are possible. MI with multivariate approaches, on the other hand, helps preserve the dependency structure by having the potential to include a large number of predictive covariates in the imputation model. To assess the validity of our imputations, we compared the imputed values produced by the existing hot deck approach and those produced by our MI approach in three ways: We assessed bivariate associations between wealth components as well as between net worth and other economic correlates, including household income, age, and education. We then assessed the performance of our newly imputed net worth variable as a prediction outcome in a multivariate regression model with covariates that include total household income, education, race/ethnicity, age, and marital status. Finally, we compared the distribution of our newly imputed net worth variable to external benchmarking data.

## 5. Results

In the 2013 PSID wave, 213 out of the 409 variables we selected were incomplete. The "apparent" missingness proportions – the share of cases without a valid value – varied between 0.01% and 99.72%, with a median value of 47.22%, and 40 variables had more than 80% missing values. However, the apparent missingness summary is misleading as it fails to distinguish meaningful missingness arising from "non-applicable" questions. Table 1, therefore, shows the apparent and true missingness proportions for all 18 wealth components. The latter values are much smaller than the former, indicating that most of the missing values arise from non-applicable cases. Figure 1 depicts the missingness patterns of the components for the 1,457 families with incomplete wealth information, showing neither a monotone nor nested structure.

**Table 1**. *Apparent missingness proportions (Apparent miss) and true missingness proportions (True miss) for the 18 wealth components in the 2013 Panel Study of Income Dynamics study.*

|  | Label | Apparent miss (%) | True miss (%) |
|---|---|---|---|
| W39B4 rel loans | Amount of loans from relatives | 98.68 | 0.39 |
| W39B3 legal | Amount of legal bills | 99.39 | 0.40 |
| W39B7 debt | Amount of other debt | 98.82 | 0.40 |
| A24_2 mortgage2 | Remaining principal mortgage 2 | 95.90 | 0.54 |
| W2B own estate | Amount of owned on other real estates | 89.25 | 0.57 |
| W11B own business | Owned amount of farm or business | 92.08 | 0.67 |
| W2A estate | Worth of other real estates | 89.48 | 0.81 |
| W39B2 med | Amount of medical bills | 89.83 | 0.81 |
| W39B1 stu loans | Amount of student loans | 75.47 | 1.20 |
| W39A credit | Amount of credit or store card debt | 66.60 | 1.31 |
| W11A business | Worth of farm or business | 92.79 | 1.39 |
| A20 house | House value | 50.40 | 1.48 |
| W16 stock | Profit if sold non-IRA stock | 89.44 | 2.01 |
| A24_1 mortgage1 | Remaining principal mortgage 1 | 66.66 | 2.28 |
| W34 bond | Profit if sold bonds/insurance | 91.11 | 2.48 |
| W22 annuity | Value of individual retirement accounts (IRA)/annuity | 81.82 | 2.52 |
| W6 vehicles | Profit if sold vehicles | 4.63 | 4.63 |
| W28 account | Amount of all accounts | 37.77 | 5.10 |

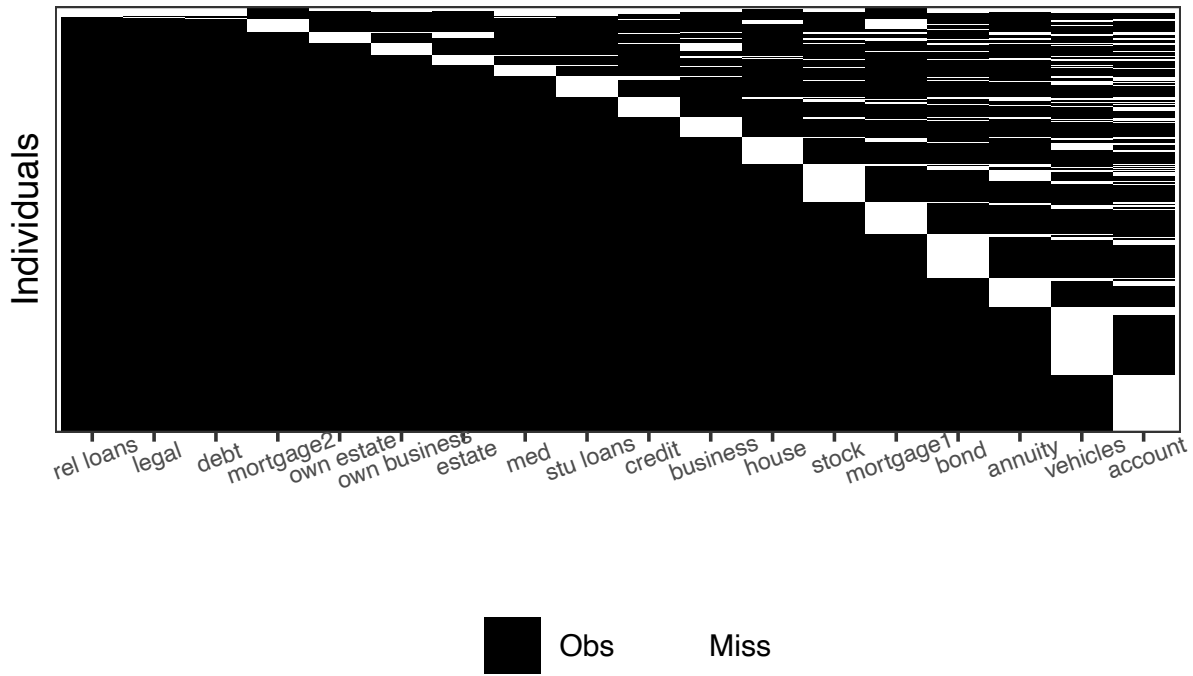

**Figure 1**: *Observed (Obs) and missing (Miss) patterns of 18 wealth components of the cases with missing wealth information in the 2013 Panel Study of Income Dynamics study.*

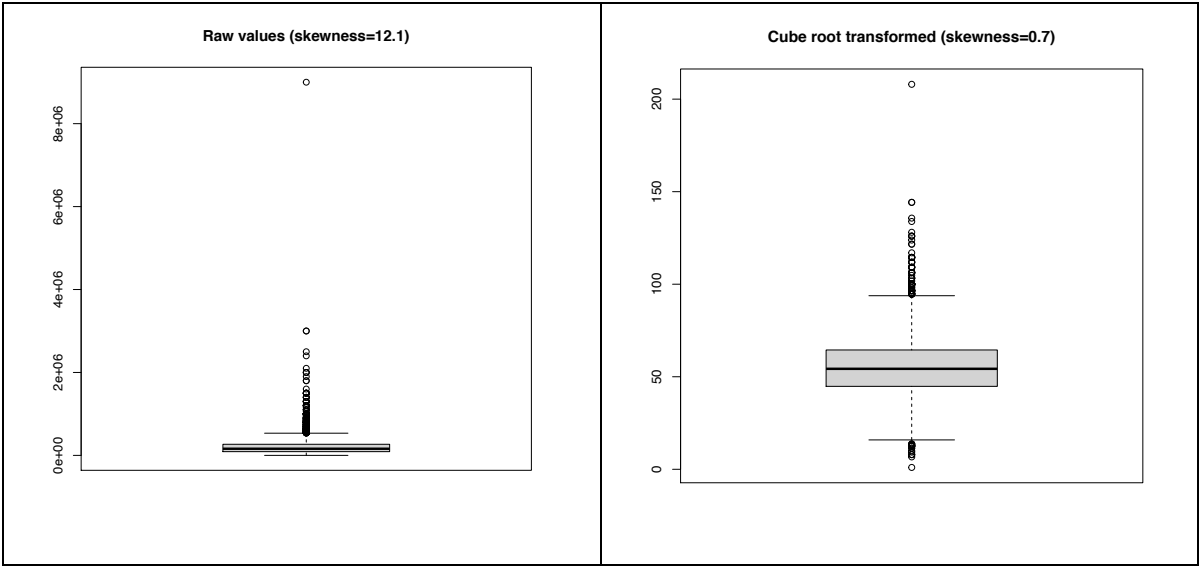

**Figure 2:** *The distribution and skewness of home values before and after the cube root transformation.*

**Table 2.** *Summary of the empirical evaluation procedures and findings in comparison of multiple imputation (MI) and hot deck (HD) imputation.*

|  | Findings |
| --- | --- |
| Summary statistics comparison between observed and imputed datasets | The summary statistics of the completed (observed and imputed data) values of MI are generally larger than those of the observed, but not dramatically so. |
| Fraction of missing information (FMI) | Most FMI values of filter and amount variables for the 18 PSID wealth components are very small. |
| Bivariate associations | MI generates larger Pearson correlation coefficient values than HD. |
| Multivariate regression | The coefficient estimates under different methods are similar. MI has smaller within-imputation variances and slightly larger overall variances than those under HD. |
| Comparison to the 2013 Survey of Consumer Finances (SCF) | The mean values from the imputation methods are similar but both are lower than the SCF estimates. The weighted estimates of business assets and other debts after MI are larger but the weighted estimates for other assets are lower than those produced by HD. |

Most of the wealth variables, especially asset/debt amounts, are severely right-skewed, with a few very large values. Figure 2 shows the frequency histograms for the original home values (restricted to the sample families that report owning a house/apartment) and after cube root transformation. The cube-root transformed values are approximately normal with a skewness of 0.7. The wealth, income, and wage variables are cube-root transformed in the imputation model.

Comparisons of results for different choices of the number of cycles suggested that about ten cycles were sufficient for most imputations, and we created $M = 10$ completed datasets. As diagnostics after imputation, Table 2 provides a summary of the procedures for empirical evaluation and the corresponding findings. Through the estimates of descriptive statistics, bivariate associations, multivariate regression models, and comparison with external data, we demonstrated the capability of MI to preserve the data dependency structure and generate plausible imputations. We found that MI reduced within-imputation variances and improved estimation efficiency over the existing single hot deck imputation method.

### 5.1. Summary statistics

Table 3 examines the summary information of total wealth and compares observed and completed datasets. The summary statistics of the completed values are generally larger than those of the observed, but not dramatically so. Table 4 presents the FMI values of filter and amount variables for the 18 PSID wealth components. Most of them are very small, primarily because of the low underlying missingness proportions. The last three columns in Table 4 compare the percentages between the observed and imputed values of positive indicators of whether the PSID household owns a wealth component and lists the number of cases with missing indicators. The comparison does not raise red flags for most components, except for the indicator A23 (Do you have a mortgage or loan on this property?) for A24_1 (Remaining principal of the first mortgage). In the imputation model for A23, the selected predictors include the house type (such as a one-family house, a two-family house, an apartment, a mobile home, or others), total family income, wage of the household head, indicator of credit card debts, and amount of the saving account. We did not have substantive concerns about the plausibility of the imputed values, given that only 23 values were missing.

**Table 3:** *Summary statistics comparison between observed (Obs) and completed (Com, the combination of observed and imputed data) 2013 PSID family wealth values based on one randomly selected multiple imputation dataset: sample size n, minimum/maximum, mean, standard deviation (Std), and quantiles. The relative difference (Rel.diff) is defined as (Com-Obs)/Obs. The wealth values are presented in $1000s and rounded to the nearest 1000.*

|       | Obs   | Com   | Rel.diff |
|-------|-------|-------|----------|
| n     | 7606  | 9063  | 0.19     |
| Min   | -995  | -995  | 0        |
| Max   | 33740 | 33740 | 0        |
| Mean  | 180   | 200   | 0.11     |
| Std   | 793   | 837   | 0.06     |
| 25th  | 0     | 0     |          |
| 50th  | 14    | 20    | 0.39     |
| 75th  | 114   | 135   | 0.18     |
| 90th  | 423   | 487   | 0.15     |
| 95th  | 846   | 912   | 0.08     |

**Table 4**: *Fraction of missing information for wealth components and the corresponding indicators, the proportions (prop) of positive indicators in the observed (obs) and imputed (imp) data based on one randomly selected imputation, and the number of cases with missing indicators.*

| Component    | FMI   | FMI of indicators | Prop. of positive indicators (obs) | Prop. of positive indicators (imp) | #missing indicators |
|--------------|-------|-------------------|------------------------------------|------------------------------------|---------------------|
| account      | 0.002 | 0.01              | 0.671                              | 0.615                              | 65                  |
| credit       | 0.006 | 0.003             | 0.344                              | 0.318                              | 44                  |
| rel loans    | 0     | 0.004             | 0.014                              | 0.061                              | 33                  |
| legal        | 0.011 | 0.003             | 0.006                              | 0                                  | 33                  |
| med          | 0.008 | 0.005             | 0.107                              | 0.182                              | 33                  |
| estate       | 0.007 | 0.002             | 0.111                              | 0.19                               | 21                  |
| own estate   | 0.012 | 0.002             | 0.111                              | 0.19                               | 21                  |
| debt         | 0.002 | 0.002             | 0.012                              | 0.03                               | 33                  |
| stu loans    | 0.014 | 0.004             | 0.255                              | 0.176                              | 34                  |
| annuity      | 0.01  | 0.002             | 0.203                              | 0.2                                | 50                  |
| business     | 0.031 | 0.001             | 0.084                              | 0.059                              | 17                  |
| own business | 0.024 | 0.001             | 0.084                              | 0.059                              | 17                  |
| bond         | 0.013 | 0.008             | 0.106                              | 0.145                              | 76                  |
| stock        | 0.012 | 0.003             | 0.12                               | 0.161                              | 56                  |
| vehicles     | 0.013 | NA                | NA                                 | NA                                 | NA                  |
| house        | 0.002 | 0                 | 0                                  | 0                                  | 0                   |
| mortgage1    | 0.023 | 0.002             | 0.355                              | 0.739                              | 23                  |
| mortgage2    | 0.031 | 0.011             | 0.043                              | 0.143                              | 35                  |

### 5.2. Bivariate associations

We first examined this MI improvement by considering the bivariate associations between wealth components. The existing imputation procedure generated the home equity value differently from MI (defined as A20 minus A24_1 minus A24_2) without reporting the three related components. This is another advantage of MI—the ability to handle hierarchical restrictions that could also be missing. Hence, we used the final home equity values, rather than the three components, and compared the resulting 16 wealth components between the two imputation methods. Figure 3 is a scatterplot of the pairwise Pearson correlation coefficients between the 16 wealth components, based on one MI dataset and the dataset imputed via the hot deck. When the correlations in the observed data are not close to 0, MI generates larger values than the hot deck method, suggesting that it is preserving associations better.

Next, we evaluated the association of the household net worth with household income, age, and education. The Bland Altman plot (Altman and Bland, 1983) in Figure 4 marks the data grouped by income quartiles and compares the imputed wealth values from the hot deck and one randomly selected MI dataset. The relationship between income and wealth is higher after MI than after hot deck imputation. As this relationship is expected to be positive (Killewald et al, 2017), this comparison favors the MI method.

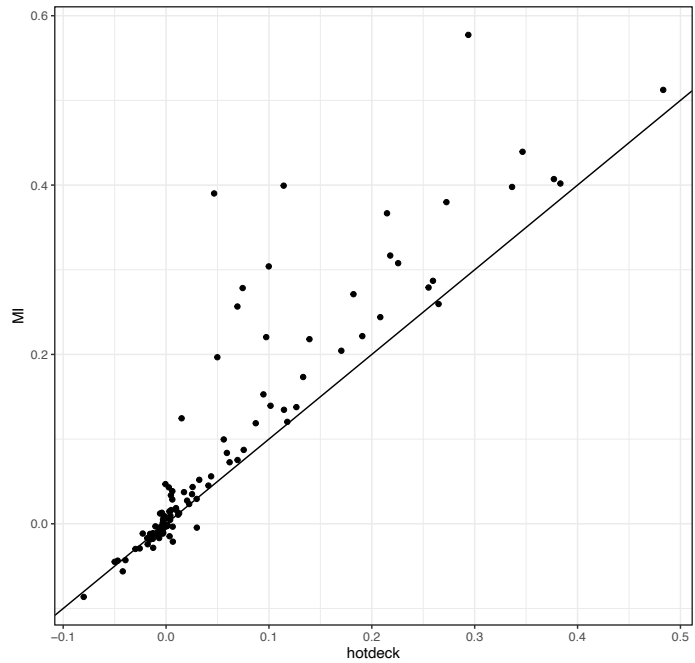

**Figure 3:** *Pairwise correlation coefficients between 16 wealth components.*

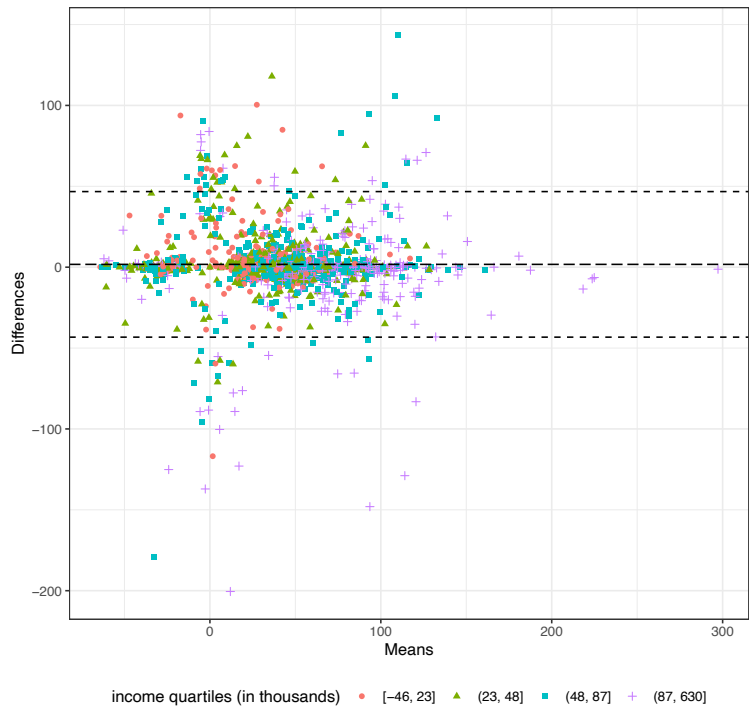

**Figure 4:** *The Bland Altman plot in the comparison of imputed wealth values (cube-root transformed) from multiple imputation and hot deck imputations. The three dashed lines represent the mean of differences minus two standard deviations, mean of differences and mean of differences plus two standard deviations. The critical difference (i.e. two times the standard deviation of differences) is 50. Some wealth values are negative because of high debts and low assets.*

**Table 5**: *Correlation between total wealth (cube-root transformed) and socio-demographics. models in comparison of observed (Obs) data, multiple imputation (MI) and hot deck (HD) imputed (imp) data.*

|  | Obs | MI-imp | HD-imp |
|---|---|---|---|
| Total family income (cube-root transformed) | 0.47 | 0.42 | 0.36 |
| Age (in years) | 0.41 | 0.37 | 0.35 |
| Education (in years) | 0.19 | 0.14 | 0.12 |

Table 5 examines the ten imputed datasets and takes the average of the Pearson correlation coefficients between total net worth and total household income, age, and education, respectively. We used the cube-root transformed values for both wealth and income to dampen the effects of large outlying values. MI preserves the dependency structure from the observed data, while the univariate hot deck imputation attenuates the correlation estimates. We also compared the regression coefficient estimates of these socio-demographic predictors in the univariate regression models with net worth as the outcome and found that the relationships in MI were larger and closer to the observed structure than those after the hot deck imputation.

Table 5 does not report the corresponding variance estimates on either the observed or imputed dataset. Because the missingness proportion is small, inferences based on the completed datasets are similar between the two imputation methods. We used Table 5 to exemplify the different imputation methods' properties for preserving relationships among the imputed variables. We checked the relationship between total household income with the 18 wealth components, respectively, and MI yields correlation coefficient estimates closer to those based on the observed values than the current hot deck method.

**5.3. Multivariate regression**

**Table 6**: *Coefficient estimates and variances (var) for the wealth regression models in comparison of observed (Obs) data, completed data (com) from multiple imputation (MI) and hot deck (HD) imputations. The sample size is n, and the number of completed datasets is M.*

|  | Obs | MI-com | HD-com | Overall var | Within var |
|---|---|---|---|---|---|
|  | (n=7606) | (n=9063, M=10) | (n=9063, M=10) | (MI/HD) | (MI/HD) |
| Intercept | -55.25 (-60.35, -50.15) | -55.33 (-60.14, -50.51) | -54.13 (-58.92, -49.34) | 1.01 | 0.99 |
| Family income | 1.07 (1.01, 1.13) | 1.1 (1.04, 1.16) | 1.07 (1.01, 1.13) | 1 | 1 |
| Education(years) | 0.53 (0.22, 0.85) | 0.56 (0.26, 0.85) | 0.57 (0.27, 0.87) | 1 | 1 |
| Non-Hisp black | -10.68 (-12.35, -9.01) | -11.57 (-13.12, -10.01) | -11.01 (-12.56, -9.47) | 1.01 | 0.99 |
| Other race/eth | -7.66 (-10.1, -5.23) | -7.47 (-9.77, -5.18) | -6.74 (-9.01, -4.47) | 1.03 | 0.98 |
| Age | 0.95 (0.89, 1.01) | 0.96 (0.9, 1.02) | 0.96 (0.89, 1.02) | 1 | 1 |
| Never married | -0.01 (-2.07, 2.05) | 0.18 (-1.8, 2.16) | 0.01 (-1.93, 1.95) | 1.04 | 0.99 |
| Widowed | -8.55 (-12.25, -4.85) | -8.17 (-11.48, -4.85) | -8.1 (-11.41, -4.79) | 1 | 0.99 |
| Divorced | -10.66 (-12.95, -8.37) | -10.63 (-12.81, -8.44) | -10.82 (-12.97, -8.68) | 1.04 | 0.99 |
| Separated | -7.58 (-11.03, -4.13) | -7.55 (-10.9, -4.2) | -7.65 (-10.94, -4.35) | 1.03 | 0.99 |

Note: The wealth and income values are cube-root transformed. The reference levels are non-Hispanic white for race/ethnicity and married for the marital status.

We fit a multivariate regression model with cube-root transformed values of household net worth as the outcome, and covariates that included total household income (cube-root transformed), education, race/ethnicity, age, and marital status, which are some of the most commonly studied correlates of a household's wealth position (Killewald et al., 2017). Table 6 shows the coefficient estimates based on the observed values as complete case analysis, and the completed datasets after MI and hot deck imputation. We applied combining rules to the ten MI datasets to propagate the missing data uncertainty. The results under the three methods are similar, likely because the missingness proportions are small. We calculated the ratio of the MI variance estimates to those from hot deck imputation, the overall variance, and the average within-imputation variances. MI includes the additional variance component accounting for missing data uncertainty, and the overall variances are slightly larger than those under hot deck imputation; but MI has smaller within-imputation variances, suggesting efficiency gains within imputation.

**Table 7:** *Mean values of total wealth (including home equities) and key wealth components across different methods and sources. MI: multiple imputation; HD: hot deck; wt: weighted; and SCF: the Survey of Consumer Finances.*

|  | HD | MI | HD–wt | MI–wt | SCF |
|---:|---:|---:|---:|---:|---:|
| Total net worth | 202058 | 200383 | 316361 | 314129 | 470491 |
| Business assets | 29264 | 29564 | 48372 | 46963 | 110128 |
| Checking/savings | 19361 | 18404 | 29637 | 28725 | 47537 |
| Stocks | 32671 | 32845 | 56759 | 57438 | 67276 |
| IRA/private annuities | 33927 | 33636 | 52452 | 52144 | 60023 |
| Net worth of vehicles | 13444 | 13405 | 14301 | 14362 | 15002 |
| Equity in primary residence | 58408 | 58480 | 86427 | 86310 | 103932 |
| Equity in real estate | 18610 | 18461 | 28561 | 28547 | 52984 |
| Other assets | 8582 | 7903 | 10335 | 10033 | 24840 |
| Other debts | 12209 | 12316 | 10484 | 10393 | 11230 |

**5.4 Comparison with external data**

In addition to the internal checks of imputation plausibility just presented, we compared our imputation results to external data. We contrasted our estimates with another study, the Survey of Consumer Finances (Federal Reserve Board, 2020). The Survey of Consumer Finances (SCF) is a triennial cross-sectional survey of U.S. families and collects information on families' balance sheets, pensions, income, and demographic characteristics. The SCF oversamples wealthy households, so we applied survey weights to the imputed datasets and generated population-representative estimates. Table 7 compares the mean values of key wealth components based on the 2013 PSID MI and hot deck imputed datasets after weighting and the weighted 2013 SCF data (Kennickell, 2000; Bricker et al., 2020). Because the PSID oversamples low-income families, the weighted PSID wealth estimates are higher than the unweighted. The values from the imputation methods are similar but both are lower than the SCF estimates as the SCF oversamples high-income families (Pfeffer et al., 2016).

Since the wealth components have right-skewed distributions, we compare the percentiles for key distribution points: the 5$^{th}$, 10$^{th}$, 25$^{th}$, 50$^{th}$, 75$^{th}$, 90$^{th}$, and 95$^{th}$ percentiles, shown in eTable 1 of the Supplement. We omitted the calculated percentiles of 0 values across all methods. Generally, the PSID estimates are lower than the SCF estimates, the latter of which have to adjust for the oversampling of wealthy families. MI generates lower values for checking/saving balances and stocks but larger values for individual retirement accounts/private annuities and other debts than the hot deck imputation. The weighted estimates of business assets and other debts after MI are larger but the weighted estimates for other assets are lower than those produced by hot deck imputation.

## 6. Conclusion and discussion

Chained-equation imputation methods are flexible and can handle general missing data patterns of missing data. Imputations can be tailored to multiple variable types. We discussed some of the methodological and practical issues encountered when we applied chained-equation MI in a large-scale, complex, particular application, the imputation of wealth data in the PSID. We demonstrated the capability of MI to preserve the data dependency structure and improve estimation efficiency. We recommend the practical implementation of MI and call for existing MI software to be extended to accommodate the methodological issues and computational hurdles with massive datasets.

Issues addressed include (a) distinguishing missing and non-applicable values, and tailoring the underlying regression models to applicable cases, (b) transformations of amount variables to reduce skewness, (c) simultaneous imputation of presence and amount of semi-continuous variables, (d) minimizing collinearity in the regressions, and (e) model checking.

The imputation models are tailored to variable types and the imputations account for restrictions and boundaries. The forward selection was applied to restrict the predictors to variables associated with the variable being imputed, speeding the convergence of the algorithm. Multiple sequences of conditional imputation models yield $M > 1$ completed datasets that can be used for standard analyses and MI inferences that propagate imputation uncertainty.

Little (2020) summarizes the important factors in this imputation setting, such as the sample size, fraction of missing information, missingness mechanism, form and strength of the true relationship between the variable with missing data and predictors, form and strength of the true relationship between missingness and its predictors, degree of association between the propensity to respond and the variable with missing values, and degree of misspecification of the true models for missingness and the survey design variables. We have described how we addressed each of these concerns in our imputations of PSID wealth data. In this application, we compared MI with the existing single hot deck imputation method. To evaluate imputation plausibility, we examined descriptive summary statistics, bivariate association studies, multivariate regression models, and compared external estimates from the SCF study.

MI incorporates highly predictive covariates into imputation models and outperforms the hot deck imputations with imputation accuracy and efficiency gains. MI preserves the associations between PSID wealth components and the relationships between the household net worth and socio-demographic factors, which is crucial to such analyses with completed data. The imputation model should be more general than the analysis model, and MI facilitates completed data analyses with general purposes. As an important infrastructure survey to study intra-/inter-generational wealth and income dynamics, the PSID study will implement MI to preserve the important data correlation structure and release multiply imputed data for public use. This will substantially reduce the random variation and facilitate wealth fluctuation studies.

The detailed solutions to practical difficulties of applying MI to the PSID study also apply to large-scale studies in general. The key properties of MI are building proper imputation models that capture dependency structure of the data and propagate the imputation uncertainty. Hence, our case study offers theoretical and practical guidance to researchers implementing MI in their own data.

This comprehensive investigation invites several extensions. First, we omitted the design information in the imputation process, and further effects are necessary on practical software development to account for design information into MI with large-scale datasets. Second, we focus on the 2013 cross-sectional PSID data. The PSID's panel structure with repeated measures of the same family across time may further improve the imputation model with highly predictive variables, such as measures from previous waves of the longitudinal study, where a two-fold MI procedure could be potentially useful (Welch et al., 2014). Third, although IVEware has a few capabilities that are not available in other MI software, further software developments are needed to fit multilevel models to account for clustering effects or correlations of longitudinal measures in imputation, specify informative prior distributions of regression coefficients to allow sophisticated variable selection, and implement flexible models or machine learning algorithms to improve the imputation performances.

Zhu, J., and Raghunathan, T. (2015), "Convergence Properties of a Sequential Regression Multiple Imputation Algorithm," *Journal of the American Statistical Association*, 110(511), 1112-1124.

**Supplemental materials for "Multiple Imputation with Massive Data: An Application to the Panel Study of Income Dynamics" by Si et al.**

**eTable 1**: *Percentiles of total wealth (including home equities) and key wealth components across different methods and sources. MI: multiple imputation; HD: hot deck; wt: weighted; and SCF: the Survey of Consumer Finances.*

| Percentiles | Hot deck | MI | Hotdeck-wt | MI-wt | SCF |
|---|---|---|---|---|---|
| Total net worth | | | | | |
| 5th | -40000 | -41141 | -32000 | -33000 | -23600 |
| 10th | -15000 | -15800 | -9100 | -9500 | -4600 |
| 25th | 0 | 0 | 2900 | 2500 | 6300 |
| 50th | 20600 | 20000 | 54000 | 53650 | 64200 |
| 75th | 138000 | 137000 | 270000 | 270000 | 226200 |
| 90th | 488000 | 486000 | 790000 | 778800 | 794900 |
| 95th | 927900 | 911534 | 1360000 | 1327683 | 1676300 |
| Business assets | | | | | |
| 90th | 0 | 0 | 0 | 0 | 1001 |
| 95th | 20000 | 20000 | 60000 | 67000 | 100082 |
| Checking/savings | | | | | |
| 10th | 0 | 0 | 0 | 0 | 20 |
| 25th | 0 | 0 | 20 | 20 | 580 |
| 50th | 1200 | 1200 | 3000 | 3000 | 3803 |
| 75th | 10000 | 10000 | 16000 | 17000 | 20016 |
| 90th | 40000 | 38000 | 60000 | 60000 | 69357 |
| 95th | 80000 | 80000 | 118000 | 116745 | 141916 |
| Stocks | | | | | |
| 90th | 5000 | 4500 | 50000 | 50000 | 30025 |
| 95th | 80000 | 80000 | 200000 | 200000 | 174142 |
| IRA/private annuities | | | | | |
| 75th | 0 | 0 | 5000 | 5000 | 7006 |
| 90th | 60000 | 60000 | 130000 | 130000 | 112092 |
| 95th | 190000 | 197233 | 300000 | 300000 | 295241 |
| Net worth of vehicles | | | | | |
| 25th | 1500 | 1500 | 2000 | 2000 | 3703 |
| 50th | 7000 | 7000 | 8000 | 8000 | 9408 |
| 75th | 18000 | 18000 | 20000 | 20000 | 18715 |
| 90th | 30000 | 30000 | 35000 | 35000 | 32927 |
| 95th | 45000 | 45000 | 50000 | 50000 | 44837 |
| Equity in primary residence | | | | | |
| 50th | 0 | 0 | 15000 | 15000 | 26021 |
| 75th | 65000 | 65000 | 105000 | 105000 | 116095 |
| 90th | 175000 | 175000 | 250000 | 250000 | 265216 |
| 95th | 275000 | 275000 | 355000 | 360000 | 417341 |
| Equity in real estate | | | | | |
| 90th | 4000 | 5000 | 40000 | 40000 | 50041 |
| 95th | 76000 | 75000 | 150000 | 150000 | 210172 |
| Other assets | | | | | |
| 75th | 0 | 0 | 0 | 0 | 1501 |
| 90th | 500 | 500 | 7000 | 6683 | 20016 |
| 95th | 20000 | 20000 | 35000 | 35000 | 60049 |
| Other debts | | | | | |
| 50th | 500 | 500 | 130 | 200 | 300 |
| 75th | 10450 | 11000 | 8000 | 8500 | 8207 |
| 90th | 34912 | 35000 | 30000 | 30000 | 28023 |
| 95th | 60000 | 60000 | 51500 | 52000 | 52443 |

**eTable2:** *Names of variables included for wealth imputation in the 2013 Panel Study of Income Dynamics data (The labels are available in https://simba.isr.umich.edu/VS/s.aspx)*

"AQCASE" "A4" "A8" "A19" "A20" "A21" "A22" "A23" "A23A_1" "A23B_1" "A24_1" "A28_1" "A23A_2" "A23B_2" "A24_2" "IMP_A20" "A20L" "A20H" "IMP_A19" "IMP_A23" "IMP_A24_1" "IMP_A28_1" "IMP_A24_2" "A24_1L" "A24_1H" "A24_2L" "A24_2H" "W1" "W1A" "W2A" "W2B" "W6" "W10" "W11A" "W11B" "W15" "W16" "W21" "W21A" "W22" "W27" "W28" "W33" "W34" "W38A" "W38BLEGL" "W38BMED" "W38BOTR" "W38BRELS" "W38BSTU" "W39A" "W39B1" "W39B2" "W39B3" "W39B4" "W39B7" "IMP_W11A" "W11AL" "W11AH" "IMP_W11B" "W11BL" "W11BH" "IMP_W28" "W28L" "W28H" "IMP_W2A" "W2AL" "W2AH" "IMP_W2B" "W2BL" "W2BH" "IMP_W16" "W16L" "W16H" "IMP_W6" "W6L" "W6H" "IMP_W34" "W34L" "W34H" "IMP_W22" "W22L" "W22H" "IMP_W39A" "W39AL" "W39AH" "IMP_W39B1" "W39B1L" "W39B1H" "IMP_W39B2" "W39B2L" "W39B2H" "IMP_W39B3" "W39B3L" "W39B3H" "IMP_W39B4" "W39B4L" "W39B4H" "IMP_W39B7" "W39B7L" "W39B7H" "IMP_W1" "IMP_W10" "IMP_W15" "IMP_W21" "IMP_W27" "IMP_W33" "IMP_W38A" "IMP_W38BSTU" "IMP_W38BMED" "IMP_W38BLEGL" "IMP_W38BRELS" "IMP_W38BOTR" "KEY_MISS" "NFU" "NKIDS" "NELDERS" "NFUHUCOLLEGE" "NOFUMCOLLEGE" "H49" "D1CKPT" "TOTHU" "GENMARSTAT" "IWLENMINS" "CCS" "G5" "G99" "G100" "G101" "G103" "G104" "G106" "G107" "G108" "G109" "G110" "G112" "G113" "G114" "WTRMETRO" "BNUM" "HBNUM" "WBNUM" "HWBIZ" "FARMY" "HDWG" "BNUS" "OVTM" "TIP" "COMS" "OTHINC" "PROF" "XTRA" "HDRENT" "HDDIV" "HDINT" "HDTRUST" "WFEARN" "WFRENT" "WFDIV" "WFINT" "WFTRUST" "TAXHW" "HDADC" "HDSSI" "HDOTRWELF" "HDVA" "HDPENS" "HDANN" "HDIRA" "HDUNEMP" "HDWRKCOMP" "HDCHSUP" "HDALIMONY" "HDHELPREL" "HDHELPNON" "HDMISC" "WFADC" "WFSSI" "WFOTRWELF" "WFPENS" "WFANN" "WFIRA" "WFUNEMP" "WFWRKCOMP" "WFCHSUP" "WFHELPREL" "WFHELPNON" "WFMISC" "OFLAB" "OFASSET" "OFADC" "OFSSI" "OFOTRWELF" "OFVA" "OFPENS" "OFUNEMP" "OFCHSUP" "OFMISC" "HDSSEC" "WFSSEC" "OFSSEC" "TOTFAMY" "FDHM13" "FDOUT13" "FDDEL13" "RENT13" "UTIL13" "TELINT13" "VEHLN13" "VEHPAY13" "VEHLS13" "AUTOIN13" "VEHADD13" "VEHREP13" "GAS13" "PARK13" "BUS13" "CAB13" "OTRAN13" "ED13" "CHILD13" "HOS13" "DOC13" "PRESCR13" "HINS13" "HHREP13" "FURN13" "CLOTH13" "TRIPS13" "OTHREC13" "FIPSTATE_DUM1" "FIPSTATE_DUM2" "FIPSTATE_DUM3" "FIPSTATE_DUM4" "FIPSTATE_DUM5" "FIPSTATE_DUM6" "FIPSTATE_DUM7" "FIPSTATE_DUM8" "FIPSTATE_DUM9" "FIPSTATE_DUM10" "FIPSTATE_DUM11" "FIPSTATE_DUM12" "FIPSTATE_DUM13" "FIPSTATE_DUM14" "FIPSTATE_DUM15" "FIPSTATE_DUM16" "FIPSTATE_DUM17" "FIPSTATE_DUM18" "FIPSTATE_DUM19" "FIPSTATE_DUM20" "FIPSTATE_DUM21" "FIPSTATE_DUM22" "FIPSTATE_DUM23" "FIPSTATE_DUM24" "FIPSTATE_DUM25" "FIPSTATE_DUM26" "FIPSTATE_DUM27" "FIPSTATE_DUM28" "FIPSTATE_DUM29" "FIPSTATE_DUM30" "FIPSTATE_DUM31" "FIPSTATE_DUM32" "FIPSTATE_DUM33" "FIPSTATE_DUM34" "FIPSTATE_DUM35" "FIPSTATE_DUM36" "FIPSTATE_DUM37" "FIPSTATE_DUM38" "FIPSTATE_DUM39" "FIPSTATE_DUM40" "FIPSTATE_DUM41" "FIPSTATE_DUM42" "FIPSTATE_DUM43" "FIPSTATE_DUM44" "FIPSTATE_DUM45" "FIPSTATE_DUM46" "FIPSTATE_DUM47" "FIPSTATE_DUM48" "FIPSTATE_DUM49" "FIPSTATE_DUM50" "FIPSTATE_DUM51" "FIPSTATE_DUM52" "FIPSTATE_DUM53" "FIPSTATE_DUM54" "FIPSTATE_DUM55" "FIPSTATE_DUM56" "FIPSTATE_DUM57" "CHGMS_DUM1" "CHGMS_DUM2" "CHGMS_DUM3" "CHGMS_DUM4" "CHGMS_DUM5" "CHGMS_DUM6" "CHGMS_DUM7" "CHGMS_DUM8" "FCC_DUM1" "FCC_DUM2" "FCC_DUM3" "FCC_DUM4" "FCC_DUM5" "FCC_DUM6" "FCC_DUM7" "FCC_DUM8" "FCC_DUM9" "FCC_DUM10" "WF_IND" "OF_IND" "SEX" "AGE" "ES" "BD1_FLAG" "BD2_FLAG" "GEOGMOBIL" "ENRLD" "GD_ENRLD" "P0P70ACKPT" "P1" "P16" "P20" "P20L" "P20H" "H1HW" "H1AHW" "H1BHW" "H1CHW" "H2HW" "H3HW" "H4HW" "H5AHWSTROK" "H5BHWHATTA" "H5CHWCORON" "H5DHWBLPRE" "H5EHWASTHM" "H5FHWLUNG" "H5GHWDIABE" "H5HHWARTH" "H5IHWMEMLO" "H5JHWLEARN" "H5KHWCANCE" "H5LHWNERVE" "H5MHWOTR" "H8HW" "H8ANITES" "COMPED" "KL39" "KL40_1" "AVGHRS" "TOTOTHRS" "WKSILLOTR" "WKSILLSLF" "WKSVAC" "WKSSTRIKE" "WKSLAYOFF" "WKSUNEMP" "WKSOOLF" "AGE_WF" "ES_WF" "BD1_FLAG_WF" "BD2_FLAG_WF" "GEOGMOBIL_WF" "ENRLD_WF" "GD_ENRLD_WF" "P0P70ACKPT_WF" "P1_WF" "P16_WF" "P20_WF" "P20L_WF" "P20H_WF" "H1HW_WF" "H1AHW_WF" "H1BHW_WF" "H1CHW_WF" "H2HW_WF" "H3HW_WF" "H4HW_WF" "H5AHWSTROK_WF" "H5BHWHATTA_WF" "H5CHWCORON_WF" "H5DHWBLPRE_WF" "H5EHWASTHM_WF" "H5FHWLUNG_WF" "H5GHWDIABE_WF" "H5HHWARTH_WF" "H5IHWMEMLO_WF" "H5JHWLEARN_WF" "H5KHWCANCE_WF" "H5LHWNERVE_WF" "H5MHWOTR_WF" "H8HW_WF" "H8ANITES_WF" "COMPED_WF" "KL39_WF" "KL40_1_WF" "AVGHRS_WF" "TOTOTHRS_WF" "WKSILLOTR_WF" "WKSILLSLF_WF" "WKSVAC_WF" "WKSSTRIKE_WF" "WKSLAYOFF_WF" "WKSUNEMP_WF" "WKSOOLF_WF"